\documentclass[10pt,amsmath,amssymb,superscriptaddress]{iopart}
\usepackage{iopams}
\usepackage{setstack}
\usepackage[dvips]{graphicx}
\usepackage{epsfig}
\usepackage{dcolumn}
\usepackage{bm}
\usepackage{longtable}

\begin{document}

\title{Metric characterization of cluster dynamics on the Sierpinski gasket}

\author{E. Agliari$^{1,2,3}$, M. Casartelli$^{1,2}$ and E. Vivo$^{4}$}
\address{$^1$ Dipartimento di Fisica, Universit\`a degli Studi di
Parma, viale Usberti 7/A, 43100 Parma, Italy}
\address{$^2$ INFN, Gruppo Collegato di
Parma, viale Usberti 7/A, 43100 Parma, Italy}
\address{$^3$ Theoretische Polymerphysik, Universit\"{a}t Freiburg, Hermann-Herder-Strasse 3, 79104 Freiburg, Germany}
\address{$^4$ Dpto. de Matem\'aticas and Grupo Interdisciplinar de Sistemas Complejos (GISC), Universidad Carlos III de Madrid, Avenida Universidad 30, 28911 Legan\'es, Spain}

\begin{abstract}
We develop and implement an algorithm for the quantitative characterization of cluster dynamics occurring on cellular automata defined on an arbitrary structure. As a prototype for such systems we  focus on the Ising model on a finite Sierpsinski Gasket, which is known to possess a complex thermodynamic behavior.
Our algorithm  requires the projection of evolving configurations into an appropriate partition space, where an information-based metrics (Rohlin distance) can be naturally defined and worked out in order to detect the changing and the stable components of clusters. The analysis highlights the existence of different temperature regimes according to the size and the rate of change of clusters. Such regimes are, in turn, related to the correlation length and the emerging ``critical'' fluctuations, in agreement with previous thermodynamic analysis, hence providing a non-trivial geometric description of the peculiar critical-like behavior exhibited by the system. Moreover, at high temperatures, we highlight the existence of different time scales controlling the evolution towards chaos.\\
\emph{Keywords}: Dynamical processes (Theory), Classical phase transitions (Theory), Classical Monte Carlo simulations
\end{abstract}


\section{\label{sec:intro}Introduction}

The effect of inhomogeneity on the critical behaviour of magnetic systems has been considered in various contexts (e.g. disorder, coupling randomness, quasiperiodic structures); in particular, discrete-spin models defined on fractal topologies possess critical properties significantly different and richer than those found for translationally invariant systems \cite{gefen1,gefen2,gefen3,milan,cassi,done}. 

The interest in fractal structures is not purely theoretical: many condensed-matter systems display strong nonuniformity on all length scales and can therefore be characterized as fractal objects; examples include the backbone of percolation clusters, aggregates obtained from diffusion-limited growth processes, and absorbent surfaces.

One of the most known fractals is the Sierpinski gasket (SG), which, due to its exact decimability, allows analytical approaches; in particular, by means of renormalization group techniques, it was proved that the Ising model on the SG exhibits phase transition only at zero temperature, while at any finite temperature the system breaks into domains and loses long-range order \cite{gefen1,gefen2,gefen3,donetti}. While this result was found in the thermodynamic limit, at the mesoscopic sizes peculiar  
and interesting thermodynamic properties arise \cite{liu}. More precisely, the Ising model defined on a finite SG exhibits critical-like
 features at nonzero, low temperatures and the solution found in the thermodynamic limit turns out to be a poor approximation. This anomalous behavior has been investigated from a thermodynamic point of view and derives from long-range, slowly decaying correlations at low temperatures \cite{luscombe}.

The way such a behavior is reflected by the evolution of spin configurations is an item so far overlooked (even if, for rectangular lattices, the idea of studying cluster dynamics may be traced back to Peierls and Griffith \cite{peierls, griffith}). Given the importance of the dynamics of Ising-like clusters in many research areas, from condensed matter to biological systems \cite{appl1,appl2,appl3}, the definition and the development of proper tools for this kind of analysis would be very useful. 
Moreover, it would be particularly intriguing for inhomogeneous substrates, due to the emerging non-trivial thermodynamic behavior; on the other hand, it is just on such structures that the definition of a proper metrization or evolutionary dynamics can be more awkward. 

The interest in cluster mobility actually extends to an extremely wide class of models: indeed the Ising model on the SG may be seen as a particular realization of cellular automata on graphs, i.e. discrete time dynamical systems assigning to each node of a graph $\mathbf G$ a value chosen in a alphabet $\mathbb K $, along a rule depending only on a finite neighborhood at previous time
\cite{wolfram}.

In this work we aim to introduce and develop proper algorithms for the quantitative characterization of cluster dynamics on graphs, and we use this approach for the Ising model on a finite SG, meant as a prototype of cellular automata on graphs. The procedure requires a projection of evolving configurations into an appropriate {\sl partition space}, where an information-based metrics (Rohlin distance) and  a method measuring the effective emergence of configurational novelty (reduction process and amplification parameter) may be naturally defined and worked out in order to focus the changing and the stable components of configurations. The  algorithmic implementations of Rohlin distance and related quantities are deeply affected by the topological features of the substrate \cite{nostro}. For instance, in previous implementations designed specifically for automata on regular lattices, the very passage from one to two dimensions yields a much higher order of complexity, and could not be exported on different substrates \cite{parti1,parti2,soc,entro}. 
On the contrary, the algorithm developed here can be directly applied to generic cellular automata for which the metric characterization of cluster dynamics gets feasible; a brief description is given in the appendix $B$.

Our investigations on the Ising SG highlight the existence of two ``critical'' temperatures, $T_I$ and $T_{II},$ demarcating three main regimes which recover, both qualitatively and quantitatively, the results of the previous thermodynamic analysis; in addition, within the above mentioned regimes, we obtain a more detailed dynamical and geometrical characterization. In particular, at very low temperatures
($T < T_I$), a long-range order is established, and the few small-sized clusters display poor overlap from one time step to the next one: this makes the distance close to zero, and the reduction ineffective. As the spot sizes start to increase, the slowness of the evolution is still such that both non-similarity and overlap between successive configurations rapidly grow.
At greater temperatures ($T_I < T < T_{II}$), large-scale correlations start to decay, clusters of all sizes appear, and overlap gets easier; then, for $T> T_{II}$, any trace of order has vanished and, even if overlaps are very important, the complexity of magnetic pattern is irreducible. The progression of such different kinds of non-similarity and dynamical overlap is well described by our parameters (distance, amplification and intersection, see below).
We will also evidence the existence of different scales controlling the disappearance of local and correlated order. 
Thus, the phenomenology provided by our method, confirms with a deeper geometric insight the peculiar critical-like behavior exhibited by the system

We finally notice that, in view of future extensions to non-equilibrium situations, we will adopt a microcanonical dynamics \cite{ACV2009}, which allows implementations even in the presence of temperature gradients.

In the next two sections, after recalling basic notations on graphs, we review some facts concerning the thermodynamic properties of the Ising model on the Sierpinski gasket (Sec.~\ref{sec:SG}), and the microcanonical dynamics working as evolutionary dynamics (Sec.~\ref{sec:Micro}). Then, we introduce general procedures for cluster identification and reduction (Sec.~\ref{sec:metrica}) and we show our results on the SG (Sec.~\ref{sec:numerics}). Finally, we present our conclusions and perspectives (Sec.~\ref{sec:con}). Technical remarks about partitions on graphs can be found in the appendices.

\section{\label{sec:SG}Ising model on the Sierpinski gasket}

A generic graph $\mathbf{G}$ is mathematically specified by the
pair $\{\Lambda, \Gamma \}$ consisting of a non-empty, countable
set of points $\Lambda$ joined pairwise by a set of links
$\Gamma$. 
The cardinality  $|\Lambda| = N$ of $\Lambda$ represents the number of sites making up the graph, i.e. its
volume. From an algebraic point of view, a graph
is completely described by its
adjacency matrix $\mathbf{A}$. Every entry of this off-diagonal, symmetric
matrix corresponds to a pair of sites, and it equals $1$ if and
only if this couple is joined by a link, otherwise it is $0$.

\begin{figure}[tb] \begin{center}
\includegraphics[width=.30\textwidth]{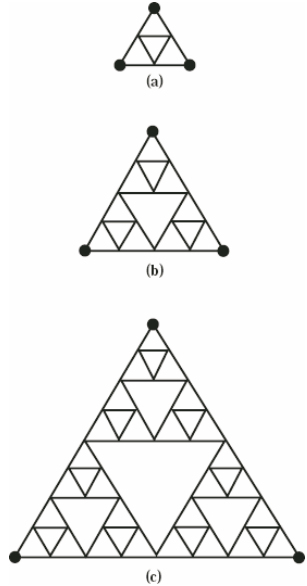}
\caption{\label{fig:sierp} (Color on line) Sierpinski gasket of generation
$2$ (a), $3$ (b) and $4$ (c), with volume $N=6$, $N=15$ and $N=42$,
respectively.}
\end{center}
\end{figure}

Here we consider the Sierpinski gasket which can be built recursively with the following procedure: the initial state ($\mathbf{G}_0$) is a triangle and the $g$th stage $\mathbf{G}_g$ is obtained joining two of the three external corners of three $\mathbf{G}_{g-1}$ to form a bigger triangle (see Fig.~\ref{fig:sierp}). In this way, the volume of $\mathbf{G}_{g}$ is $3(3^g-1)/2$. The Gasket is obtained as the limit for $g \rightarrow \infty$ of this procedure.

\subsection{\label{ssec:termo_SG}Thermodynamic properties of the Ising model on the SG}
The Ising model on a generic graph $\mathbf{G}=\{\Lambda, \Gamma \}$ is defined associating the spin variable $s_i = \pm 1$ to every site $i$ of the graph, and considering a nearest-neighbours interaction between points $i$ and $j$, such that $A_{ij}=1$. The Hamiltonian is therefore
\begin{equation}
H_{\mathbf{G}}(\mathbf{s},J) = - J \sum_{i=1}^N \sum_{j>i}^N A_{ij} s_i s_j,
\end{equation}
where $\mathbf{s}=\{s_i\}_{i \in \Lambda}$ denotes the magnetic configuration of the system and the coupling $J$ is assumed to be the same for any couple; in the following we will set $J \equiv 1$.

As it is well known, a magnetic model defined on a finite lattice cannot exhibit critical behaviour at nonzero temperatures; critical features can only emerge when the underlying lattice becomes infinitely large, i.e. in the thermodynamic limit. Another necessary condition in order to have a nonzero critical temperature concerns the topology of the (infinite) substrate: for Euclidean structures it was rigorously shown that the dimension must be larger than 1. Analogously, it has been shown that the discrete symmetry Ising model on finitely ramified fractals cannot have a nonzero critical temperature \cite{gefen3,luscombe,bhattacharya}.
However, the critical behaviour of one-dimensional systems and finitely ramified fractals can be markedly different, since for the latter it is further governed by additional geometric aspects such as ramification, lacunarity, and connectivity \cite{gefen1}.

The thermodynamic properties of the Ising model on the SG were studied in details in \cite{gefen1,gefen2,gefen3,milan,liu,luscombe,stosic}, where it was shown that its scale-invariant, fractal structure leads to highly cooperative correlations and, at sufficiently low temperatures, the correlation length $\xi$ becomes extremely large and slowly decaying (compared with the one-dimensional case), so that any system with size smaller than $\xi$ displays long-range order. 
Indeed, one can define an apparent magnetic transition temperature $T_1$ as the point where $\langle m^2 \rangle =0.5$; for a system of generation $g$ ($g \gg 1$),
$T_1 \approx 4J / \ln(4g)$, for example, for generation $g=5$ and $g=6$ one finds $T_1 \approx 1.335$ and $T_1 \approx 1.259$, respectively. 
Hence, as the system size is enlarged, $T_1$ diminishes slowly; more precisely, being $N=3(3^g+1)/2$ the number of nodes for a gasket of generation $g$, one has $T_1 \sim 1/ \ln[\ln(N)]$.

On the other hand, the specific heat $c$ does not display any anomaly associated with long-range order, yet it exhibits a peak at a temperature $T_2 \approx 2J$  \cite{liu}. Conversely, the ``reduced'' specific heat $\sim ~ c T^2$, (basically the derivative of the magnetic energy per link with respect to the inverse temperature \cite{luscombe}), evidences a qualitative difference between the SG and a one-dimensional system, since for the former it exhibits a peak, while for latter it grows monotonically as the temperature is increased. Indeed, for the SG, as $T$ is increased from small values, the energy decreases rather slowly as a result of a relatively large cost due to the large (bulk) coordination number; at larger temperatures it becomes progressively easier to reduce the energy and this reflects the fact that the large fluctuations begin to develop; finally, since the energy must ultimately vanish, the rate of change gets smaller and smaller \cite{luscombe}.

In conclusion, the SG displays a non-trivial thermodynamic behavior which can be summarized as follows. 
At $T< T_1$ a long-range order is established; for $T_1 < T < T_2$ a short-range order is still present with large  fluctuations on all length scales. At larger temperatures $T>T_2$ any trace of order has disappeared and a paramagnetic state is approached.

\section{\label{sec:Micro}Microcanonical Dynamics}
When studying transport properties, microcanonical dynamics are usually chosen as they allow to describe an isolated system, or its isolated bulk, without any assumption on
the equilibrium state between the system and the surrounding.

Here we adopt a recently introduced microcanonical dynamics \cite{ACV2009}, which features a high degree of flexibility, being ergodic in any temperature range and implementable on a generic structure, even in the presence of disorder. On regular lattices (e.g. cylinder, torus) such a dynamics has already been shown to be able to lead the system to thermalized states compatible with those expected from a canonical dynamics and to allow the study of out-of-equilibrium properties \cite{ACV2009,future}.   

Although we will focus only on equilibrium regimes, the reason for choosing this dynamics is twofold: First we test its reliability on an inhomogeneous structure; second, we pave the ground for the study of transport properties on such a substrate.

In the following, we briefly resume how it works having in mind as substrate a generic graph $\mathbf{G}$ described by the adjacency matrix $\mathbf{A}$.

For each pair of connected sites, namely each link $i, j$, such that $A_{ij}=1$, besides the magnetic energy $E^m_{ij} = J s_i s_j$, we introduce a local {\it kinetic energy} $E_{ij}>0$ which is, in principle, unbounded.
Now, the dynamical rule proceeds as follows:
\begin{enumerate}
\item Start from a (discrete) distribution of energies $\{E_{ij}\}_{i,j=1,...,N}$;
\item choose randomly a link  $i \sim j$;
\item extract one over the possible four spin-configurations for the couple of sites $i,j$, and evaluate the magnetic energy variation $\Delta E^m_{ij}$ induced by the move;
\item if $\Delta E_{ij}^m \leq 0$, accept the move and increase the link energy $E_{ij}$ of $\Delta E_{ij}^m$. When $\Delta E^m_{ij} >0$, the move is accepted only if $E_{ij} \geq \Delta E_{ij}^m$ and the link energy is consequently decreased of $\Delta E_{ij}^m$; otherwise  the move is not accepted and the link energy is not updated.
\end{enumerate}

It is worth remarking that $\Delta E_{ij}^m$ allows for energy variations occurred on the link  $i \sim j$ as well as on those pertaining to links adjacent to sites $i$ or $j$:
$$
\Delta E_{ij}^m = 2 s_i \delta_i  \sum_{k \in \Lambda} A_{ik} s_k + 2 s_j \delta_j  \sum_{k \in \Lambda} A_{jk} s_k - 4 s_i s_j \delta_i \delta_j,
$$
where $\delta_i = 1$ if the $i$th spin has undergone a spin flip, otherwise it is zero.
It is therefore clear that, due to the discreteness of the system and to the fact that $J$ is constant over all links, both $E^m_{ij}$ and $E_{ij}$ are discrete variables: the former can only assume two different values corresponding to the aligned and non aligned configurations of the adjacent spins $i$ and $j$; the latter can only assume integer values deriving by proper combinations of the pertaining $\Delta E_{ij}^m$.

We also notice that $E_{ij}$ works as an additional degree of freedom and the above dynamics conserves the total energy given by the following Hamiltonian function
\begin{equation}\label{ham}
H_{\mathbf{G}}(\mathbf{s},\{ E_{ij} \})=\sum_{i,j \in \Lambda}A_{ij}  \left( s_i s_j + E_{ij}\right).
\end{equation}

As shown in \cite{ACV2009}, the magnetic and kinetic energies result to be non-correlated: this allows a natural definition of temperature at equilibrium, which, in a very natural way, depends only on the average kinetic energy. In fact, the link energy satisfies the Boltzmann distribution $\exp(-\beta E_{ij})$, and the fitted constant $\beta=1/T$ just corresponds to the expected inverse temperature of the system. 

Finally, we stress that the possible coupling with thermostats set at a temperature $\bar{T}$ can be realized straightforwardly by selecting a subset of links $\Gamma' \subset \Gamma$ (or, analogously a subset of nodes) and by extracting the pertaining kinetic energies according to the Boltzmann distribution, being $\beta = 1/ \bar{T}$  \cite{ACV2009}.

\section{Metrization}\label{sec:metrica}
As foresaid, an interesting characterization of the Ising model on the SG may be accomplished by a configurational analysis defined in the wider context of Cellular Automata on graphs.
Such an analysis can be realized by a particular metrization referring non directly to the configuration space, whose points are the states $\mathbf a, \mathbf b,\mathbf c,.. $ of the system, but to a peculiar {\sl partition space} containing - among other elements - the cluster distributions of the system. The mathematical framework, which  is summarized in the Appendix $A.1$, requires that the graph is endowed with the structure of a probability space. Precisely, we consider the triple $(\mathbf G,\mathcal G, \mu ) $, where the measure $\mu$ on the subset algebra $\mathcal G $ of the gasket $\mathbf G$ is simply given by the normalized number of nodes in every subset. A {\sl configuration} (or {\sl state}) on $\mathbf G$ is a function assigning to each node a value in an alphabet
$\mathbb K $. The set of all possible configurations will be denoted as $\mathcal C (\mathbf G)$, the configuration space.
Since we consider the Ising model on $\mathbf G $, the alphabet is binary, but all we are going to say is independent of the number  $ \mid \mathbb K \mid $. The adjacency matrix, combined with the list of values on nodes, allows an easy definition
of clusters on an arbitrary structure (the procedure is identical to the recognization of connected subset in graph colouring): two homogeneous nodes belong to the same cluster if they are connected through a path of nodes sharing the same value. Thus, every state $\mathbf a$ determines in a natural way  a {\sl partition}  of $\mathbf G $, i.e. an exhaustive collection $\alpha = \Phi(\mathbf a )  \equiv \{ A_1, A_2,...,A_n \} $ of disjoint subsets $ A_i$, each connected and homogeneous, commonly called {\sl atoms} of the partition. The set of all partitions of $\mathbf G$ constitutes the {\sl partition space} $\mathcal Z \equiv  \mathcal Z (\mathbf G )$. The application $ \Phi : \mathcal C (\mathbf G) \to \mathcal Z (\mathbf G )$ is many-to-one, since, for instance, permutations in $\mathbb K $ produce different states but the same partition. Obviously, $\mathcal Z (\mathbf G ) $  contains much more partitions than those derived from clusterization, e.g. since, in general, atoms do not require to be connected sets, as clusters are. In the present case,  $\mathcal Z (\mathbf G)$ is discrete and finite because $\mathbf G$ is such, but the formalism applies in abstract probability spaces (see \cite{bill,AA,rohlin,sinai,rohlin2}).

Clearly, when a dynamics is defined on the graph, this determines configuration orbits  $\{ \mathbf a(t_n) \} $ starting from any initial state $ \mathbf a(t_0) $, and the corresponding partition orbits $\{ \alpha(t_n) \}$, where $ \alpha(t_n)  \equiv  \Phi (\mathbf a(t_n) )$

Basic operations between two arbitrary partitions $\alpha$ and $\beta$  are the minimal common multiple 
$ \gamma = \alpha \vee  \beta$, and the maximal common factor (m.c.f.) or {\it intersection} $\sigma = \alpha \wedge \beta$  (see Fig.~\ref{fig:example} and Appendix $A.1$ for details).
The entropy $H({\sigma})$ of the intersection $\sigma$, when calculated between partitions at next steps along an orbit, is an index of the relevance of the (instantaneous) non evolving part.

The metrization of the partition spaces is  based on the Rohlin distance, which describes the non-similarity of two arbitrary partitions $\alpha$ and $\beta$. This distance, requiring the Shannon conditional entropy $H(\alpha | \beta) $ of measurable partitions (see  Eq.~\ref{condiz}), is given by the functional of Eq.~\ref{rohlin}, we anticipate here:
\begin{equation}
d_R (\alpha,\beta)
=H(\alpha|\beta)+H(\beta|\alpha).
\end{equation}
In order to amplify non similarity, in the Appendix $A.2$ we present also a method, referred to as ``reduction process'' and denoted $\pi$, which acts on couples of partitions and uses both operations $\vee$ and $\wedge$. More precisely, given two partitions, say $\alpha$ and $\beta$, their reduction is obtained by first defining their intersection $\sigma = \alpha \wedge \beta$ and by keeping from both partitions only those subfactors $ \widehat{\alpha}_k$ and $ \widehat{\beta}_k$, prime with $\sigma$, namely such that $\widehat{\alpha}_k \wedge \sigma = \widehat{\beta}_k \wedge \sigma = \nu$. Then, the reduced partitions are given by, $\widehat{\alpha} = \vee_j \widehat{\alpha}_j$ and $\widehat{\beta} = \vee_j \widehat{\beta}_j$ respectively (see Fig.~\ref{fig:All}, lower panel). The process $\pi$ gives evidence of the essentially different sub-partitions of any couple in $ \mathcal Z (\mathbf G )$, and therefore amplifies their distance: $d_R(\alpha ,\beta) \leq d_R(\pi(\alpha ,\beta))$. Hence, by comparing the distance between reduced and non-reduced couples, it is possible to introduce an {\sl amplification ratio}, that is $d_R( \pi[\alpha, \beta])/d_R( \alpha, \beta)$, which provides further information about the cluster distribution and mobility. However, all this analysis has to be performed in correlation with other observables. As explained in Appendix $A.3$, the reduction process is effective, namely gives rise to a large amplification ratio, whenever one of the two partitions, say $\alpha$, displays at least one cluster which in $\beta$ is exactly decomposed into smaller ones. Now, in the case under study $\alpha$ and $\beta$ are partitions defined by cluster configurations at two successive steps, and the existence of a common cluster is furthered by the special topology of the SG: clusters corresponding to subgraphs which are (combinations of) gaskets of generation $<g$ are rather stable (the border is made by two vertices only), nonetheless internal fluctuations may occur and hence decompose the cluster itself.
  
It is worth underlining that while the    $ \gamma = \alpha \vee  \beta$ operation is rather trivial, the $\sigma = \alpha \wedge \beta$ operation and the reduction process $\pi$ are quite tricky. Reduction in particular is the main algorithmical obstacle in handling large graphs ($g > 6$).

Of course, the Rohlin distance is deeply different from the well known Hamming distance $d_H$ in the configuration space, i.e. 
\begin{equation} \label{eq:dH}
d_H (\mathbf a,\mathbf b) = {1\over N}\sum_1^{N} \rho (a_i,b_i),
\end{equation}
where $a_i$ and $b_i$ are the values of the $i-$th node, $\rho$ is a distance functional in the alphabet. The simplest distance in $ \mathbb K $ is $\rho(x,y)= 1 - \delta_{xy}$, leading to Hamming distance 0.5 for purely random configurations. Looking for instance to Fig.~\ref{fig:All} (upper panel), we would get the maximal Hamming distance for the configurations $\mathbf{a}$ and $\mathbf{b}$,  and a minor distance for  $\mathbf{a}$ and $\mathbf{c}$,  while the Rohlin metrics on the corresponding partitions gives null distance in the former case and a high distance in the latter. 
In other terms, $d_R$ and $d_H$ are  deeply different in principle
not only because they refer to different objects (defined in  $\mathcal Z(\mathbf G) $ and $\mathcal C (\mathbf G)$, respectively), but also because the former has do to with mutual distribution of clusters, which could involve geometrical features and long range correlations, while the latter is the sum of strictly local differences. Therefore, even if the correspondence $\Phi : \mathcal C \to \mathcal Z $ is many-to-one, the loss of information should be compensated by the fact that by $d_R$ we get a global estimate on cluster distributions, instead of a bare sum of uncorrelated differences.
Finally, more details concerning the implementation of the algorithm for measure of the above mentioned metric observables can be found in Appendix $A.4$.

\section{Numerical Experiments and Results}\label{sec:numerics}

We now focus on the Ising model on the SG, with the microcanonical dynamics described in Sec.~\ref{sec:Micro}. 
The equilibrium regime is ensured  by coupling the system to thermostats set at the same  temperature; in this particular case the simplest way is to consider, for a given generation $g$, the external triangle, i.e the $3 \times 2^{g-1}$ links defining the perimeter of the gasket; if then we exclude the six angular links, we get $3$ separated thermostats. 
This could be useful in the future, as it consistutes a quite simple way of imposing temperature gradients to the system. However, it should be noticed that, due to the fact that the contact between the thermostats and the system gets vanishingly small with respect to the bulk as $g$ is increased, the time for thermalization is expected to grow with the size of the system. Anyway, as mentioned, we are now interested only in the equilibrium behavior.

Before proceeding, we underline that our checks strongly confirm that the thermalized states reached by the microcanonical dynamics are consistent with those expected from a canonical dynamics, e.g. based on the Metropolis algorithm. In particular, we verified that, for a given temperature, macroscopic observables like the magnetization and the energy measured with the two kinds of dynamics are indistinguishable.  

For any given temperature $T$, the geometric observables have been calculated as time series starting after a thermalization time, and lasting an observation time $t_{\mathrm{max}}$, where time is measured in units of Monte Carlo (MC) steps; i.e. $N$ elementary moves.  
In particular, we consider finite segments of trajectories $\{ \mathbf a(t_k) \} $ in $\mathcal C$ or $\{ \alpha(t_k) \equiv \Phi (\mathbf a(t_k)  )\} $ in $\mathcal Z$, for $t_k = 0,1,2,...,t_{\mathrm{max}}$, as well as the related
intersections  $\sigma(t_k) = \alpha(t_k) \wedge \alpha(t_{k+1})$ and the couples of reduced partitions at successive time steps
$(\widehat \alpha ,\widehat \beta ) = \pi(\alpha ,\beta ) $.
From such equilibrium trajectories we obtain segments of time series for the following  quantities:

\begin{enumerate}
\item the entropy $H(\cdot)$, applied either to the orbits $\{ \alpha(t_k)\}$ or $\{ \sigma(t_k)\}$;
\item the Rohlin distance $d_R(t_k) \equiv  d_R( \alpha(t_{k-1}), \alpha(t_k))$, (see Eq.~\ref{rohlin});
\item the amplified Rohlin distance $ \widehat{d_R}(t_k) \equiv d_R( \pi[\alpha(t_{k-1}), \alpha(t_k)])$;
\item the amplification ratio $\widehat{d_R}(t_k)  / d_R(t_k)$;
\item the Hamming distance $d_H(t_k) = d_H(\mathbf{a}(t_{k-1}),\mathbf{a}(t_k))$, (see Eqs.~\ref{eq:dH} and \ref{hamming}).
\end{enumerate}

Here we focus to the case of unitary time steps ($t_{k+1}- t_k$ equals one MC step), leaving the study of the role of the time gap to future investigations.

After setting the experimental parameters (size, thermalization time, confidence length of trajectories $t_{\mathrm{max}}$), we calculate time averages, variances etc. for each of the series above. Results found for different choices of size (we especially focused on gaskets of generation $g=4$, $g=5$ and $g=6$, corresponding to $N=120 $, $N=366$ and $N=1095$ sites respectively) and of $t_{\mathrm{max}}$ are qualitatively in very good agreement. Moreover, to approach the thermalized state, the initial configuration is taken ferromagnetic; this minimizes the likelihood of pinning effects during the evolution \cite{ale}. In the following we report only the essential information, dropping redundant numerical outputs.

In general, the observables analyzed highlight the existence of different regimes, demarcated by remarkable temperatures $T_I\approx 1.3$ and $T_{II} \approx 2.0$. More precisely, $T_I$ corresponds to 
the flex in the Rohlin distance and to the peak in the amplification ratio, while $T_{II}$ corresponds to the peak in the variance of Rohlin distance and to the crossover in the intersection entropy (see Figs.~$2-4$). Interestingly, $T_I$ and  $T_{II}$ recover the ``critical'' temperatures $T_1$ and $T_2$ 
evidenced by thermodynamic analysis (see Sec.~$2$ and \cite{gefen1,gefen2,liu}). Indeed, consistently with thermodynamic results, the highlighted regimes correspond to a long-range order region and to a disordered  region with a critical-like transition region in between. More precisely:  

\begin{figure}\label{fig:d_r}
\includegraphics[width=1.0\textwidth]{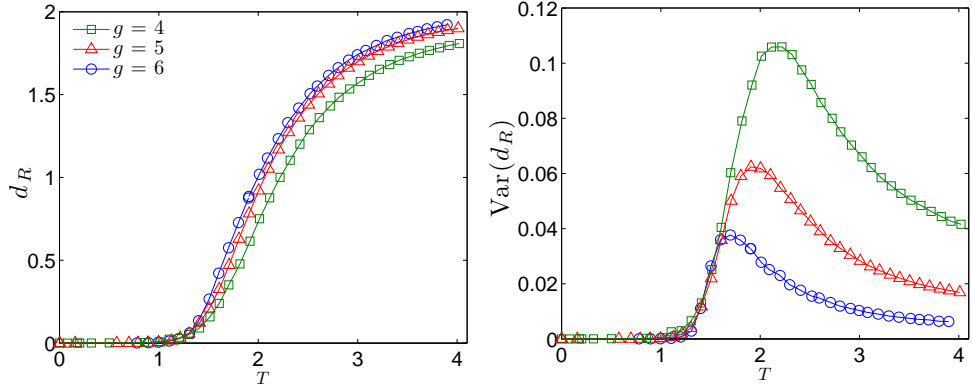}
\caption{(Color on line) Mean (left panel)  and variance (right panel) for Rohlin's distance $d_R$ as a function of the temperature; three different sizes are compared.}
\end{figure}

\begin{figure}\label{fig:d_h}
\includegraphics[width=1.0\textwidth]{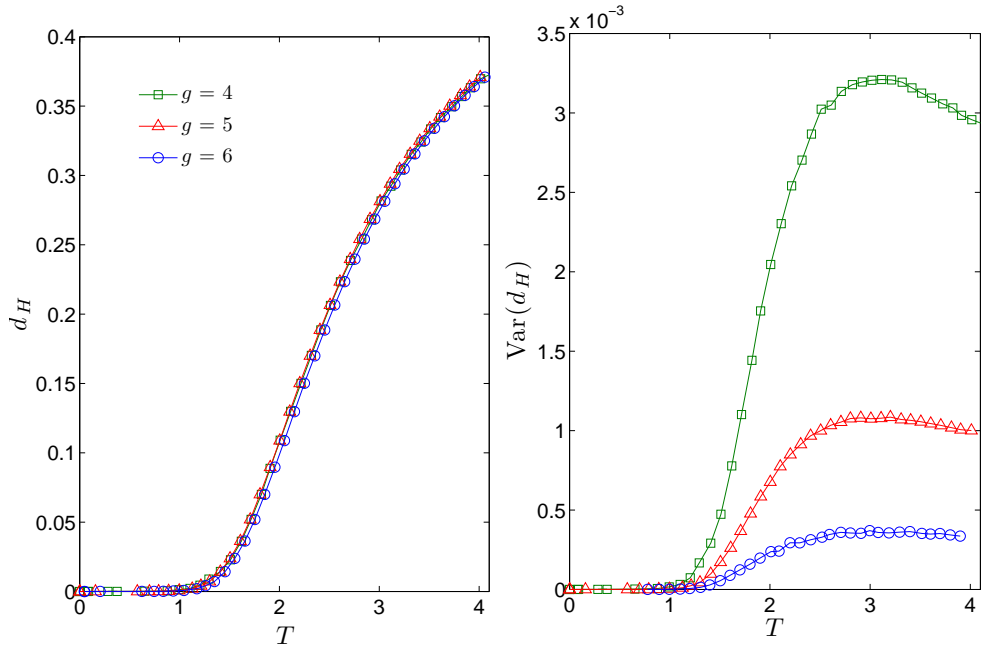}
\caption{(Color on line) Mean (left panel)  and variance (right panel) for Hamming's distance $d_H$ as a function of the temperature; three different sizes are compared.}
\end{figure}

\begin{itemize}
\item for $ T < T_{I} $, $\langle m^2 \rangle$ is close to $1$ (see \cite{liu}), while the distances $d_H$ and $d_R$ are approximately $0$ (see Figs.~$2$ and $3$); in fact, at such small temperatures a ferromagnetic order is established over large length-scales: clusters are constituted by few single spots in the large ``sea'' of equally oriented spins and  spin-flips are rather unlikely to happen. So, configurations - and partitions - at consecutive steps
differ for such small spots that distances are  extremely small. Moreover, during a MC step, the rare spin flips occurring are yet able to change the atoms in such a way that overlaps between consecutive partitions are quite improbable. This inhibits the reduction, and the amplification is close to 1. 
For $T$ approaching $T_I$,  such spots get larger, but, due to the slowness of the evolution at low temperature, their borders can remain sufficiently unchanged for several steps. Hence, the emergence of spots within such clusters allows the reduction, processing couples of consecutive partitions, to get more effective, and the time average of the amplification is manifestly increased. The growth of the spots can be retrieved by the intersection entropy increase, as shown in the inset of Fig.~$4$ (left panel). 

\item for $T \approx T_I$, $\langle m^2 \rangle$ is close to $0.5$ and both $d_H$ and $d_R$ start to be significantly larger than $0$. This  is a consequence of the fact that spin flips are getting more frequent. More interestingly, the amplification ratio reaches a maximum. 
We have seen that, at the middle of the previous regime, two conditions cooperated to start the growth of the amplification ratio: non empty intersection $\sigma$, and the fact that at consecutive steps there exist uniform large clusters which are decomposed internally, yielding an effective reduction (as explained in Appendix $A.3$). Here, the peak in the amplification ratio means that cluster sizes and the speed of the dynamics optimally fit such conditions; moreover, fixed clusters are more likely. The exact determination of the peak temperature is tricky, due to the growing complexity of the configuration dynamics, but we argue that $T_I$ 
(or $T_1$) is a good approximation. This behavior is consistent with the apparent transition occurring at $T_1$ as a long-range order breakdown.

\item for $T_I < T < T_{II}$,  $\langle m^2 \rangle$ is approaching zero. Clusters get more intricate due to the growing temperature, and  $H({\sigma})$, which measures the relevance of overlapping between successive configurations, exhibits a rapid growth (see Fig.~$4$, left panel). The coexistence of fragmentation  of large clusters, which is a signature of decay for long range correlations, and overlapping may be related to ``critical slowing down'' effects \cite{CSD}.  Clearly, fragmented clusters have a higher probability to overlap, but are unlikely to include small spots; As a consequence the amplification ratio is still larger than $1$, though rapidly decreasing. 
Note that $H({\sigma})$ is rather  far from saturation (namely, uncorrelated chaos) and this suggests the persistence of  
a local short-length order. 
\item for $ T \sim T_{II}$, the amplification ratio is practically 1: such a breakdown of the reduction, as forementioned,  has a completely different meaning than in the case of very low temperatures, where overlapping was insignificant for the smallness of
sparse clusters in the ``big sea'' of dominant magnetic orientation. Here, due to the smallness of the $\sigma$ atoms, it is unlikely that a sufficiently large cluster is internally decomposed at the next step (see also the Appendix $A.3$). 

At this temperature the variance of $d_R$ has a maximum (whose value scales like $1/\sqrt{N}$), and this constitutes a signature that, along the trajectory, the fluctuation  of the distance is particularly important. Indeed, the complexity of clusters shape and their mobility can give rise to wide fluctuations of distances in time. 
Also, $H({\sigma})$ exhibits a crossover: from $T_{II}$, the growth is due only to fragmentation, while the ``critical
slowing down'' is over. Both the persistent growth of $H({\sigma})$ and the value of $d_H $ (neatly below $0.5$) indicate that the complete chaos is far from being established at this
temperature and for these sizes. Moreover, the large fluctuations make the system more susceptible to spin-flips, i.e. energy changes, and this is consistent with the peak in the specific heat.

\item for $T > T_{II}$ , the Rohlin variance decreases roughly as $\sim 1/N$ toward an asymptotic value corresponding to the uncorrelated chaos, whose onset may be recognized by $d_H \simeq 0.5$. We argue that the fragmentation and the mobility of clusters stabilize the behavior of the time series for global quantities like distances.

\begin{figure}\label{fig:ampl}
\includegraphics[width=1.0\textwidth]{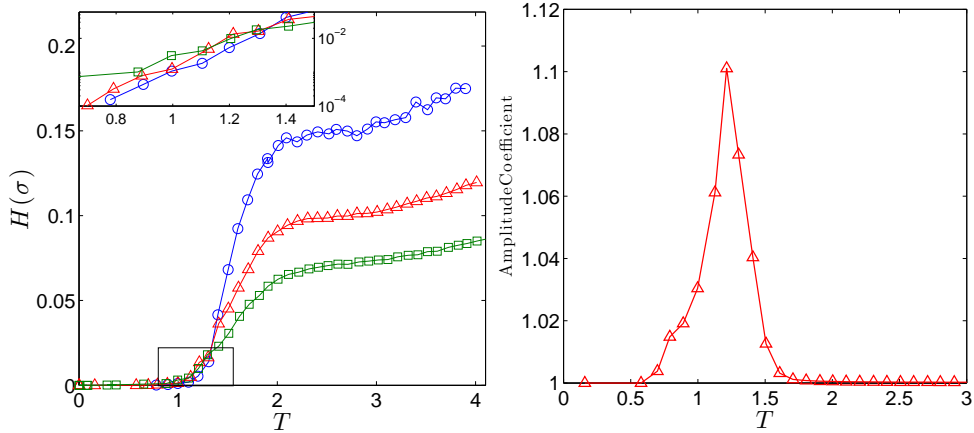}
\caption{(Color on line) Intersection entropy $H(\sigma)$ (left panel) and amplification ratio $\widetilde{d_R}/d_R$ (right panel) as a function of the temperature. For the latter, accurate numerical data are presented only for generation $g=5$, due to time consuming calculations; nonetheless we have checked that for other values of $g$ the qualitative behavior is robust. The inset of the left panel shows a magnification of the departure from zero of the entropy; notice the semilogarithmic scale.}
\end{figure}

We note however that the maximum for the $d_H$-variance occurs at $T\approx 3$, well above the corresponding maximum of the
$d_R$-variance at $T\approx 2$. We deduce that in this temperature interval big clusters of all sizes, typical of the previous fractal configurations, have started their fragmentation, so that the global shape diversity between subsequent configurations is 
 stabilizing and this explains why the $d_R$-variance decreases. However, the fragmentation process is still slow enough that the local matching between different or equal spins at next steps is highly unstable in time, allowing for the growth of the $d_H$-variance. Indeed, spins belonging to the inner part of a cluster (even if of small size)  result from the dynamics to be more stable than those belonging to its periphery. For $T> 3$ the fragmentation is such that both local overlaps and medium length correlations tend to stabilize the time behavior. There is therefore a temperature interval exhibiting a subtle interplay between correlation length and time stability. In other terms, in the way to the chaos we recognize two time scales: one in terms of global similitude (shape) of clusters along the orbits, which starts to stabilize just at the end of the ``critical slowing down''
($T\approx 2$); the other in terms of local overlaps, whose stabilization requires a higher temperature  ($T\approx 3$). The direct visual inspection could be awkward for such properties, undetectable also from the mean magnetization.

\end{itemize} 

\section{Conclusions and perspectives}\label{sec:con}

In this work we have performed a quantitative characterization of cluster dynamics for the Ising model defined on a finite Sierspinski Gasket, whose thermodynamic behavior is known to be non-trivial.
The analysis is based on a set of geometric observables, such as Hamming and Rohlin distances, and on reduction operations among partitions which allow to detect the evolving and the stable components of clusters.

The phenomenology evidenced by previous thermodynamic analysis is qualitatively and quantitatively confirmed by the present metric approach, which provides in addition a geometric characterization of the anomalous behavior of the system. Indeed, we highlight first the existence of two ``critical'' temperatures: $T_I$ corresponding to a peak in the amplification ratio, meaning that
there an abrupt change in the clusters behavior occurs, and $T_{II}$, corresponding to a peak in 
the Rohlin variance and other crossovers, evidencing the loss of short range order.  
More precisely, the amplification ratio gives detailed information about the development of clusters, starting
from little independent spots (amplification ratio equal to 1) to first overlaps (amplification ratio larger than 1 and growing to
the maximum) while the next phase of decreasing ratio indicates the complex effect of decreasing efficiency of the reduction
due to the fragmentation, up to $T_{II}$ when the fragmentation is such that,
notewithstanding the large intersection, the reduction process is inhibited by the extreme improbability of clusters with a decomposable inner part (see  \ref{ampli}).

Another new information we obtain in the high temperature regime is the existence of  different scales in the destruction of local and correlated order, as evidenced by the Hamming and Rohlin variances, with an anticipated peak for the latter quantity.

Thus, the analysis above enabled us to distinguish different behaviors and phases demarcated by $T_I$ and $T_{II}$ as main milestones; we cannot speak of course of ``phase transitions'', since the onset of these distinct behaviors seems to be a rather smooth process. However, from such a detailed description of the collective motion we get an insight on the interplay between dynamic and statistical features, especially the decay of spatial and temporal correlations.

This work opens a lot of possible extensions and further insights. 
The next step is the metric characterization of clusters dynamics in condition of non-equilibrium; 
a kind of analysis allowed by the chosen dynamics even in the presence of a disordered coupling pattern \cite{ACV2009}. 

In particular, it could be interesting to deepen the effect of the double scale in the approach to uncorrelated chaos on the conductivity. In the same context, another interesting item to explore is the relevance of topology for the cluster diffusion. Indeed, the whole set of operations performed on the SG, from cluster identification to  reduction, immediately applies to any automaton on arbitrary, connected graphs: the process depends only on the adjacency matrix, on a generic alphabet and on a dynamics, working as a proper external engine generating a succession of configurations. 
Therefore, by the bare substitution of the adjacency matrix, the algorithm is ready to fit a great variety of statistical models (e.g. Pott's model on lattices of arbitrary dimension or graphs), or other network problems where a dynamics is defined. If necessary, nodes could also be weighted, defining alternative probability measures. 
True algorithmic problems and non trivial extensions could only arise from alternative definitions of partitions (assuming e.g. atoms of a different kind with respect to the clusters) or from a different factorization, modifying the reduction process.

\appendix
\section{Technical remarks on metrization}
\subsection{Generalities}\label{general}

The formalism and general results for partition spaces and Rohlin metrics may be recovered e.g. in \cite{bill,AA,rohlin,sinai,rohlin2}. Let $(\mathbf{M}, \mathcal{M},\mu ) $ be a probability space, 
that is an arbitrary set $\mathbf{M}$, a $\sigma $-algebra $\mathcal{M}$ of subsets of $\mathbf{M}$, and a
normalized measure $\mu $ on  $\mathcal{M}$. In our case the set $\mathbf{M}$ is just given by $ \mathbf G$.

A partition of $\mathbf{M}$ is a finite collection $\alpha \equiv (A_1,A_2,...,A_n) $ of measurable
disjoint subsets covering $\mathbf{M}$, i.e. $A_i\cap A_k = \varnothing  $ and
$ \cup_k A_k = \mathbf{M}$. The $\{A_k\}$'s are called the ``atoms'' of $ \alpha $.
The set of all finite measurable partitions is denoted $\mathcal{Z} \equiv  \mathcal{Z} (\mathbf{M}) $.
The unit partition $\nu$ consists of the single atom $\mathbf{M}$.
A partial order in $\mathcal{Z}$, i.e. a relation $\alpha \leq \beta$, means that $\beta$
is a refinement of $\alpha$; equivalently, every $A_k$ is exactly composed with some  $B_j$ included in $\beta$.
In such case, $\alpha$ is said to be a ``factor'' of $\beta$. Clearly,
$\nu \leq \alpha$ for every $\alpha$.

Such  terms as ``unit'' and ``factor'' depend on a commutative and associative
pseudo-product, or composition, $ \gamma = \alpha \vee \beta$,
denoting the less refined of all partitions greater or equal to both $\alpha$ and $\beta$, whose atoms are the non empty
intersections of the $\alpha$ and $\beta$ atoms. If not ambiguous, we can also write $\gamma = \alpha \beta$.
Obviously, $\alpha \eta = \alpha$ whenever $\eta \leq \alpha$,
and in particular $\alpha \nu = \alpha$ for every $\alpha$. Such properties
make the result of this operation a kind of ``minimal common multiple''. 

\begin{figure}[tb] \begin{center}
\includegraphics[width=.80\textwidth]{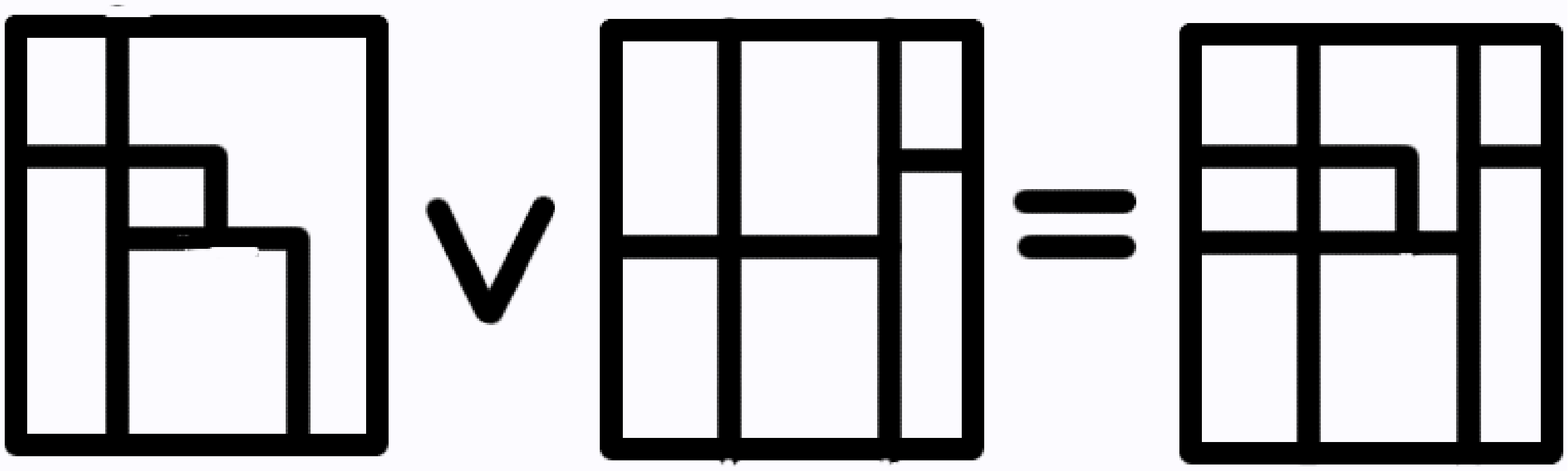}
\includegraphics[width=.80\textwidth]{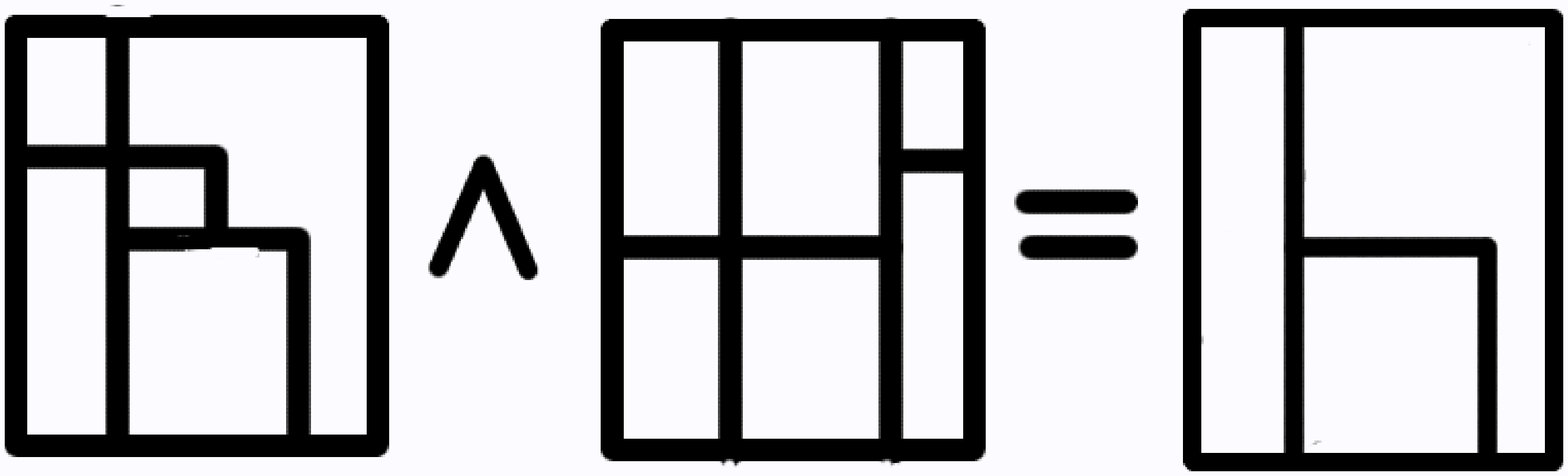}
\caption{\label{fig:example} Product (upper panel) and intersection 
(lower panel) between two examples of partition realized on a square lattice.}
\end{center}
\end{figure}

Conversely,  $\sigma = \alpha \wedge \beta $ is the greatest partition such that
$\sigma \leq \alpha $ and $\sigma \leq \beta $. In this case,
$\alpha \wedge \nu  = \nu ~$ for every $\alpha $, and $\alpha \wedge \beta  = \nu $
implies that $\alpha $ and $\beta ~$  are ``relatively prime''
(i.e. they have no common factor). Therefore, the result is
a sort of ``maximal common factor''. 
See Fig.~\ref{fig:example} for an example of product and intersection among partitions.

A partition may represent a probabilistic experiment with non overlapping outcomes
$A_1,..,A_N$, where the ``atomic'' event $A_k$ has probability $\mu(A_k)$. A factor
is therefore a sub-experiment of the finer experiment, grouping several outcomes as equivalent: for
instance, ``odd or even'' is a two-atoms sub-experiment of the $\{1,2,3,4,5,6\}$ dice experiment.

The Shannon's Entropy $H(\alpha)$ defined on every partition as
\begin{equation}
\label{shannon}
 H(\alpha)= -\sum_{i=1}^n \mu(A_i)\ln \mu(A_i) ~,
\end{equation}
is a measure of the mean information obtained from the experiment.
If  $\beta=(B_1,...,B_m)$ is
another partition, the conditional entropy of $\alpha$ with
respect to $\beta$ is
\begin{equation}
\label{condiz}
 H(\alpha|\beta) = -\sum_{i=1}^n\sum_{k=1}^m \mu(A_i\cap B_k)\ln\frac{\mu(A_i\cap
 B_k)}{\mu(B_k)}~,
\end{equation}
where, as usual, one takes $x\ln x =0$ for $x=0$. This conditional entropy is the mean residual
information obtained from  $ \alpha $ when the result of $\beta  $ is known.
Note that the Shannon entropy depends only on the distribution of the atom measures,
not on their nature or ``shape'' (this term coud have no meaning
in abstract spaces). On the contrary, the mutual relations among atoms
(and possibly their shapes) directly influence the conditional entropy
(see Figure A.2, upper row).

Now, the Rohlin distance $d_R$ is defined in $ \mathcal{Z} (\mathbf{M}) $ by
\begin{equation}
\label{rohlin} d_R (\alpha,\beta)
=H(\alpha|\beta)+H(\beta|\alpha),
\end{equation}
and it may be considered as a measure of the overall non-similarity between
$\alpha  $ and $ \beta $.

If $\mathbf M $ is finite,  a {\sl configuration} or {\sl state} $\mathbf a $ on $\mathbf M $ 
is a function assigning to each point $ x_i \in \mathbf M$ a value $ a_i = f(x_i)$ in an alphabet $ \mathbb K$. 
All possible configurations form a space $\mathcal C \equiv  \mathcal C (\mathbf M ) $. 
In $\mathcal C$ the Hamming distance  $d_H$ is defined by
\begin{equation} \label{hamming} 
d_H  (\mathbf a,\mathbf b)={\mathcal N}  \sum_i \rho (a_i , b_i ),
\end{equation}
where $ \rho (a_i , b_i ) $ is a distance in $ \mathbb K$ and $\mathcal{N}$ a possible normalization coefficient.

To each configuration corresponds an exhaustive collection of  clusters, 
i.e. connected subsets of $ \mathbf M$ with homogeneous value in $\mathbb K$, defining a particular partition 
in  $\mathcal Z (\mathbf M ) $.
This establishes a many-to-one correspondence $\Phi : \mathcal C \to \mathcal Z$, making possible
the comparison between $  d_H  (\mathbf a,\mathbf b)$ in $\mathcal C$ and $  d_R (\alpha,\beta)$ in
$\mathcal Z$, where $\alpha = \Phi(\mathbf a)$ and  $\beta = \Phi(\mathbf b)$.

\begin{figure}[tb] 
\begin{center}
\includegraphics[width=.90\textwidth]{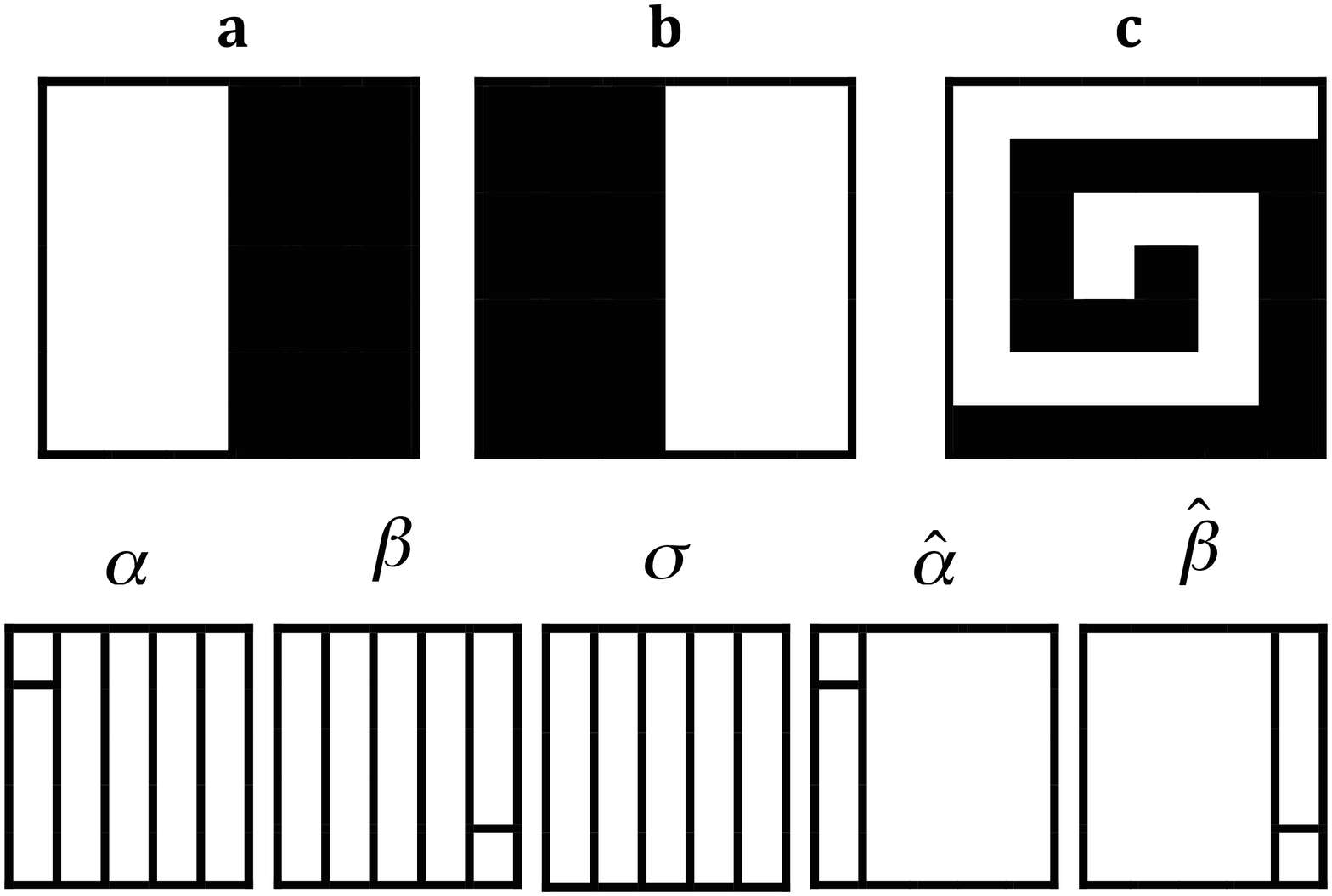}
\caption{\label{fig:All} Upper row: The three configurations depicted evidence the difference between Hamming and Rohlin distances: $d_H$ is maximal for configurations $\mathbf{a}$ and $\mathbf{b}  $, and minor for $\mathbf{a}$ and $ \mathbf{c}$; 
$d_R$ is null in the former case and large in the latter. Lower row: For the partitions  $\alpha$ and $\beta$ the common factor is given by 
$\sigma$, while $\widehat{\alpha}$ and $\widehat{\beta}$  are the reduced partitions respectively}
\end{center}
\end{figure}

\vskip 30.0 pt

\begin{figure}[tb] 
\begin{center}
\includegraphics[width=.90\textwidth]{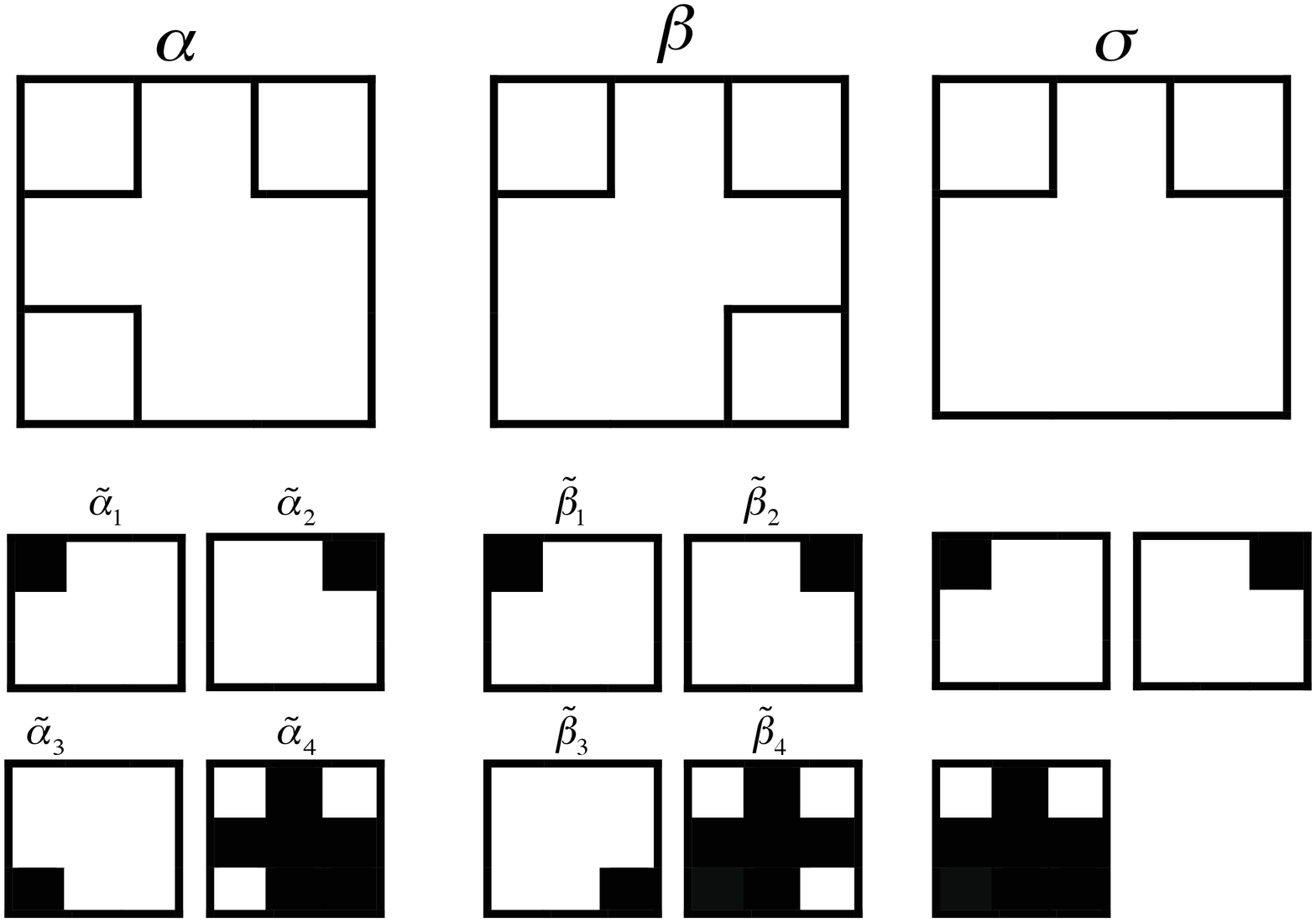}
\caption{\label{fig:Redaa}  Upper row: $\alpha$ and $\beta$ are two partitions of the square, each of four atoms; $\sigma$ is their intersection or m.c.f. of their atoms. Lower row: List of relevant elementary dichotomic factors, where the black specifies the atom and the white the complementary set. }
\end{center}
\end{figure}

\begin{figure}[tb] 
\begin{center}
\includegraphics[width=.90\textwidth]{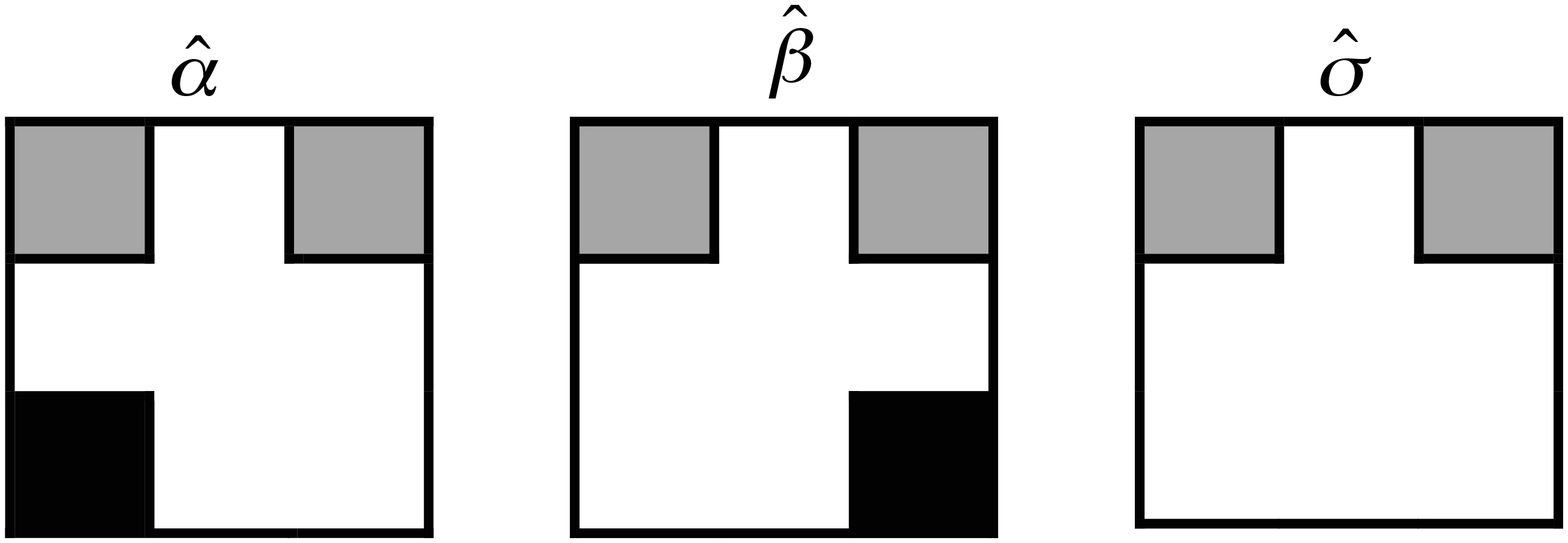}
\caption{\label{fig:Redbb}  The partitions $\hat{\alpha}$ and $\hat{\beta}$ are the reduced partitions of those appearing in figure \ref{fig:Redaa}: atoms are now three, individuated
by black, white and grey. The partition $\hat{\sigma}$ is the reduced intersection. Notice that the grey atom is non connected.}
\end{center}
\end{figure}

\subsection{Reduction }
\label{reduc}
\vskip 10.0 pt

The essential non-similarity between  two partitions could be confused and weakened
by the presence of a tight common factor, i.e. a common sub-partition (see Fig.~\ref{fig:All}, lower panel).
Therefore, we would eliminate common factors as far as possible,
a ``reduction''  which is expected to amplify the distance. However, this operation
(analogous to the reduction to minimal terms for fractions \cite{parti1}) is not
uniquely definite because partitions, differently from integers, do not admit a unique
factorization into primes. The role of primes (i.e. indecomposable) factors can
be played by dichotomic sub-partitions, which are still extremely redundant ($2^{n-1}-1 $
indeed for a partition with $n$ atoms).

For  $\alpha \equiv (A_1,A_2,...,A_n) $ we shall define therefore a restricted
family $\mathbf{E} (\alpha ) $ of ``elementary'' dichotomic factors
$ \widetilde {\alpha}_1, \widetilde {\alpha}_2, ...,\widetilde {\alpha}_n $
such that
\begin{enumerate}
\item  $\mathbf{E} (\alpha ) $ must be well defined for every  $ \alpha \in \mathcal{Z} $;
\item   $\mathbf{E} (\alpha ) $ does not contain more than $n$ (the number of atoms in $\alpha  $)
elementary factors;
\item  $\vee_{k=1}^n   \widetilde {\alpha}_k = \alpha $.
\end{enumerate}

A universal choice consists in taking as dichotomic factors $\widetilde \alpha_k \equiv (A_k,A_k^c)  $,
the partitions formed by single atoms and their complements to  $ \mathbf{M}$.
Elementary factors of this form, used  throughout in the present paper, will be called ``simple''. 

For a couple $\alpha  $ and $\beta  $, once their elementary factors   $\mathbf{E} (\alpha ) $ and $\mathbf{E} (\beta ) $ have been defined, the reduction process consists in the following steps:
\begin{enumerate}
\item  define the maximal common factor $\sigma = \alpha \wedge \beta $;
\item  drop from   $\mathbf{E} (\alpha ) $ and $\mathbf{E} (\beta ) $ those factors which are not relatively prime with $\sigma $, 
and note the surviving factors $ \widehat \alpha _k $ and $ \widehat \beta _j$ respectively 
(this means $ \widehat \alpha _k \wedge \sigma  = \widehat \beta _j \wedge \sigma  = \nu $);
\item  define $ \widehat \alpha  = \vee _k \widehat \alpha _k$ and  $ \widehat \beta  = \vee _j \widehat \beta _j$.
\end{enumerate}
The reason of step $ii$, which seems to be cumbersome with respect to the simple dropping of
common factors in   $\mathbf{E} (\alpha ) $ and $\mathbf{E} (\beta ) $, is that in  general
two families could have no common factors and, nevertheless, $\sigma \not=\nu $.
This happens, for instance, when $\alpha <\beta $ with no common elementary factors. Then $\sigma =\alpha  $ and $\widehat \alpha =\nu ~$ with the reduction above, while $\widehat \alpha =\alpha  $ with the dropping of common factors.

It results that
$d_R(\widehat \alpha ,\widehat \beta )\geq  d_R(\alpha ,\beta ) $, as requested \cite{parti1,parti2,bill}.

Note however that while two relatively prime partitions are already reduced, not necessarily two reduced
partitions are relatively prime: see for instance the case of Figure A.3, which could represent small portions of 
wider partitions deduced from almost chaotic configurations. In the next subsection we give more precise details on this. 

The correspondence $\pi ~:~(\alpha ,\beta ) \rightarrow  (\widehat \alpha ,\widehat \beta )$
is many-to-one and idempotent, i.e. $\pi \circ \pi =\pi  $. It is a sort of projection from
$\mathcal{Z} \times  \mathcal{Z}$ on the subset of irreducible pairs.

The reduction process, therefore, essentially depends on the choice of the family
$\mathbf{E} (\alpha ) $ of elementary factors. 
Besides simple factors, other families exist, which in particular cases
could prove more convenient for algorithmic reasons or for the observer's attitude
in the probabilistic experiment. While simple dichotomic factors correspond
to looking at the ``occurrence or not'' of  single atomic outcomes, other attitudes
could isolate dichotomic outcomes enjoying supplementary properties
which are not universal but depend, typically, on some additional geometrical structure of $ \mathbf{M}$ (order, connection, orientation etc.).
For instance, in previous works on rectangular lattices \cite{soc,entro} elementary factors were identified 
by external contours of clusters, a choice intended to optimize the simple connection of the factor atoms. 
This is not convenient on general graphs, where $\mathbf M \equiv \mathbf G $,  
because the determination of external surfaces could be cumbersome or impossible. 
Therefore, in the present work, elementary factors will always be the simple factors $\{A_i, A_i^c \}$.

\subsection{Reduction vs. Amplification}
\label{ampli}
\vskip 10.0 pt

Let $\alpha$ and $\beta$ such that $\sigma= \alpha \wedge \beta \not= \nu$, and $\sigma = (S_1, S_2, ... ,S_q)$. Every $S_k \in \sigma$ is the union of 
some subsets $\{A_{ki} \}$ and $\{B_{kj} \}$ of atoms in $\alpha$ and $\beta$. Possibly, such subsets may be of one single atom or more: we indicate $s^{\alpha}_k$ and $m^{\alpha}_k$ the single or multiple atoms cases for $\alpha$, and analogously 
$s^{\beta}_k$ and $m^{\beta}_k$ for $\beta$. Clearly, at fixed $k$, $s^{\alpha}_k  = A_k $ and $ s^{\beta}_k = B_k$, but it is useful to keep a distinct notation in order to remember that such atoms are not only in the partitions but also in their m.c.f.  Therefore, $S_k$ may be composed in four forms: 
$(s^{\alpha}_k, s^{\beta}_k) $, $(s^{\alpha}_k, m^{\beta}_k) $, $(m^{\alpha}_k, s^{\beta}_k)$  and $(m^{\alpha}_k , m^{\beta}_k)$. 
For instance, in Figure \ref{fig:All}, the m.c.f. $\sigma$ has one $(m^{\alpha},s^{\beta})$, one $(s^{\alpha},m^{\beta})$ and three $(s^{\alpha},s^{\beta})$ atoms, while, in Figure \ref{fig:Redaa}, $\sigma$ has one $(m^{\alpha},m^{\beta})$ and two $(s^{\alpha},s^{\beta})$ atoms.

\smallskip 

{\noindent \bf Proposition 1:} the atoms of $\widehat{\alpha}$ are all those of $\alpha$ contained in the $\{ m^{\alpha}_k\}$ groups, plus one atom constituted
by the intersection of their complementary sets, or equivalently by $ \cup_k s^{\alpha}_k$ (and similarly for  $\widehat{\beta}$).

The proof immediately follows from the fact that the  $(A_k,A_k^c)$ elementary factors are dropped in the reduction process if and only if $A_k \equiv s^{\alpha}_k$. This is true in any abstract partition space, whenever simple factors are used; notice however that, speaking of partitions generated by {\sl connected} configurations, as those considered in the present paper, this last atom could be  non-connected (an example in the figure \ref{fig:Redbb}).

\smallskip 

{\noindent \bf Proposition 2:} the atoms of ${\widehat{\sigma}} = {\widehat{\alpha}}\wedge {\widehat{\beta}} $ are:  {\sl i)} all 
the $S_k \in \sigma$ of the $(m^{\alpha}_k , m^{\beta}_k) $ form; {\sl ii)} one more atom (the complementary part to the union 
of the previous ones) if at least one term $(s^{\alpha}_k, s^{\beta}_k) $ exists; or {\sl iii)} two more atoms if, besides
the $(m^{\alpha}_k , m^{\beta}_k) $, only mixed terms  $(s^{\alpha}_k, m^{\beta}_k) $ and  $(m^{\alpha}_k, s^{\beta}_k)$ exist.

The point  {\sl i)} follows from Proposition 1, since the $\{m^{\alpha}_k \} $ and $ \{m^{\beta}_k\} $ groups reconstitute common subpartitions in $\widehat{\alpha}$ and $\widehat{\beta}$; point {\sl iii)}  depends on the fact that, in absence of $(s^{\alpha}_k, s^{\beta}_k) $ terms, 
there is an exact correspondence between 
$\cup_k  s^{\alpha}_k$ and  $\cup_k  m^{\beta}_k$ (and equivalently $\cup_k  s^{\beta}_k$ and  $\cup_k  m^{\alpha}_k$ ), where the
$k$ index runs over the mixed-type $S_k$'s only; the two supplemetary atoms are therefore  $\cup_k s^{\alpha}_k $ and $\cup_k  s^{\beta}_k$; as to point   {\sl ii)}, we observe that the complementary set of the $\{ m^{\alpha}_k \} $, i.e. 
$\cup_k s^{\alpha}_k$, cannot be exactly decomposed by the atoms of $ \{ m^{\beta}_k \} $, and vice versa, just because the presence of the supplementary  simple terms.

Several easy corollaries follow. For instance, when simple factors are used, Propositions 1 and 2 constitute a direct constructive proof that the reduction $\pi$ is a projection. Or that there is no reduction at all when in $\sigma$ there are only  $(m^{\alpha}_k , m^{\beta}_k)$ terms, or when $\sigma$ has only two atoms. Or else that in order to have ${\widehat{\sigma}}=\nu$, no terms $(m^{\alpha}_k , m^{\beta}_k)$ are allowed and at least one term $(s^{\alpha}_k, s^{\beta}_k) $ is necessary.  Moreover, if for one partition, say $\alpha$, no $\{m^{\alpha}_k \}$  group exists, then $\alpha = \sigma$ and  $\widehat{\alpha} = \nu$. Etc. 

An important item is the relation between reduction (meant as pattern simplification, as stated in Proposition 1) and the  amplification ratio  $d_R({\widehat{\alpha}},{\widehat{\beta}}) / d_R(\alpha ,\beta)$, measuring the metric effectiveness of the reduction.  

\smallskip

{\noindent \bf Proposition 3:} The necessary and sufficient conditions in order to have 
$d_R({\widehat{\alpha}},{\widehat{\beta}}) >  d_R(\alpha ,\beta)$, and therefore amplification ratio $> 1$, are the existence in
$\sigma$ of at least one atom of mixed form,  $(s^{\alpha}_k, m^{\beta}_k) $ or $(m^{\alpha}_k, s^{\beta}_k)$, and the existence of at least two distinct group $s^{x}_k$ from the same partition ($x= \alpha$ or $\beta$). 

This may be proved using the important equality
\begin{equation}\label{dR} d_R (\alpha,\beta)  = 2 H (\alpha \vee \beta) - H(\alpha) - H(\beta),
\end{equation}
(see \cite{parti1,bill}). In  the computation of all entropies appearing in Eq.~\ref{dR}, one can split the Shannon's sums (See Eq.~\ref{shannon}) along
the atoms $S_1, S_2, ..., S_q$ of $\sigma$, because the intersections $A_i \cap B_j$ appearing in $\alpha \vee \beta $ are surely
empty for different $S_k$'s. Therefore, using Propositions 1, for atoms  of the $(m^{\alpha}_k , m^{\beta}_k)$ form, the partial contributes to entropies in Eq.~\ref{dR} are the same in $  d_R(\alpha ,\beta)$ and $d_R({\widehat{\alpha}},{\widehat{\beta}})$. In absence of mixed terms, the remaining part of both $\alpha$ and $\beta$ is constituted by the same atoms 
$ s^{\alpha}_k = s^{\beta}_k$ , and the contribute to the distance is 0, exactly as for the supplementary single atom  $ \cup_k s^{\alpha}_k \equiv \cup_k s^{\beta}_k $ common to  ${\widehat{\alpha}}$ and ${\widehat{\beta}}$. 

Assume now that also mixed terms $(s^{\alpha}_k, m^{\beta}_k) $ and $(m^{\alpha}_k, s^{\beta}_k)$ are present (and possibly also terms $(s^{\alpha}_k, s^{\beta}_k) $) in such a way that at least two  $ s^{\alpha}_k$ or two $ s^{\beta}_k$ appear. Using again Eq.~\ref{dR} for these components,  $H(\alpha \vee \beta )$ and $H({\widehat{\alpha}}\vee  {\widehat{ \beta}} )$ are equal, because the intersection of the multiple atoms in one partition give the same result with the single atoms in  the other partition (before reduction) or with their union (after reduction).

As to the subtracted quantities in Eq.~\ref{dR}, the contributes from $H(\alpha)$ and $H(\beta)$ are strictly greater than $H({\widehat{\alpha}})$
and $ H({\widehat{ \beta}})$ because there are contributes from separate atoms in the former case, and from their union in the latter (this is why at least two
 $ s^{\alpha}_k$ or two $ s^{\beta}_k$ are requested). Clearly, elementary convexity properties of the  $-x\ln x$ function used here, see e.g. \cite{bill}. Therefore $d_R({\widehat{\alpha}},{\widehat{\beta}}) $ is strictly greater than $ d_R(\alpha ,\beta)$, and  after the subtraction in Eq.~\ref{dR} the distance is increased, i.e. the amplification ratio is greater than one.
 
Proposition 3 clarifies that the metric effectiveness of the reduction, besides the pattern simplification, is due to the difference between $-\sum_k  \mu (s^x_k) \ln \mu(s^x_k) \ $ and  $- \mu (\cup_k s^x_k) \ln \mu (\cup_k s^x_k) $, where $x$ is $\alpha$ or $\beta$. In particular, if $\alpha$ and $\beta$ are partitions defined by cluster configurations at two successive steps as those considered in the present paper, then the amplification requires that there exist big clusters at one time exactly decomposed at the next (or previous) time into smaller ones.

\section{The algorithm}
\label{algo}
\vskip 10.0 pt

The $N\times N$ adjacency matrix $\mathbf{A}$ of the graph  $\mathbf G$ and its state  ${\mathbf a}=(a_1,a_2,...,a_N) $, i.e. the $ \mathbb{K}$-valued list of sites labelled $(1,2,...,N )$, constitute the essential information necessary to work out the partition algebra and the metrization algorithm, which is independent of the way the states are generated (dynamics). 

By standard ``colouring'' techniques, connected sites are iteratively recognized testing their value in $ \mathbb{K}$, and the partition is easily produced as a list of grouped labels, one group for every atom. For a ten sites graph, for instance, a particular partition could be $\alpha \equiv  \{A_1,A_2,A_3\}=\{ (1,3,4),(2,10),(5,6,7,8,9) \}$, with measures $3/10,~2/10,~5/10$ for $A_1,A_2,A_3$ respectively. A  useful representation for the atoms is a binary string of length $N$: e.g., in the example above, $A_1= (1 0 1 1 0 0 0 0 0 0)$, etc.  Note that from now on the topological nature of the graph does not influence the operations, which regards only the label lists.
The ``simple factors'' of the family $\mathbf{E} (\alpha )$ are immediately defined. In the example, $ \widetilde{\alpha}_1 = \{ (1,3,4),(2,5,6,7,8,9,10) \}$, etc. (the corresponding binary strings are obviously complementary). Clearly, the extreme simplicity of this procedure could be replicated with other choices of dichotomic factors, e.g.  by taking  the ``internal-close'' and the ``external-open'' parts of Jordan contours (provided that such contours are well defined). Only in the very special case of one-dimensional chains the latter choice (internal-external, i.e. left-right) proves not only more effective but easier than the former, because contours are left extremities of semi-open intervals.

The  $\vee $ operation may be  easily implemented by boolean intersections on the atom lists. This is enough for the Rohlin distance computation by formula \ref{shannon}, thanks to the equality of Eq.~\ref{dR} above. 

More attention is required for the $\wedge$ operation, which is the key  step to define the maximal common factor $\sigma = \alpha \wedge \beta$, but the task may equally accomplished by boolean operators: every atom of $\sigma$ is built site by site, testing the simultaneous belonging to some atoms of $\alpha$ and $\beta$, up to exhaustion.
 Once the maximal common factor is defined, most of the computation time is spent in the storage and  management of the surviving simple factors $\widehat{\alpha}_k $ and $ \widehat{\beta}_j $, along the criteria of the subsection \ref{ampli}. However, all this does not imply conceptually new operations. 
Reduced partitions, and their atoms, which in principle are defined by intersections on the atoms of such surviving factors,  can take advantage of Proposition 1 in subsection \ref{ampli}, completing the $\pi$ process. For the amplified distance $\widehat{d_R}$, the functional in Eq.~\ref{dR}  is applied to the couple  $(\widehat{\alpha}_k ,\widehat{\beta}_j )$, and the observations of the previous subsection reveal useful again.

\section*{Acknowledgments}
This work is partly supported by the FIRB grant: $RBFR08EKEV$

\end{document}